\documentclass[aip,jap,amsmath,amssymb,reprint]{revtex4-1}
\usepackage{graphicx}

\begin{document}

\newcommand{\IFA}{Department of Physics and Astronomy, Aarhus
  University, Ny Munkegade 120, DK-8000 Aarhus C, Denmark.}
\newcommand{\iNANO}{Interdisciplinary Nanoscience Center (iNANO),
  Aarhus University, Gustav Wieds Vej 14, DK-8000 Aarhus C, Denmark.}

\title{Light emission from silicon with tin-containing
  nanocrystals}

\author{S{\o}ren Roesgaard}
\affiliation{\iNANO}
\author{Jacques Chevallier}
\affiliation{\iNANO}
\affiliation{\IFA}
\author{Peter I.~Gaiduk}
\affiliation{Belarussian State University, Praspyekt Nyezalyezhnastsi 
4, 220030 Minsk, Belarus.}
\author{John Lundsgaard Hansen}
\affiliation{\iNANO}
\affiliation{\IFA}
\author{Pia Bomholt Jensen}
\affiliation{\iNANO}
\affiliation{\IFA}
\author{Arne Nylandsted Larsen}
\affiliation{\iNANO}
\affiliation{\IFA}
\author{Axel Svane}
\affiliation{\IFA}
\author{Peter Balling}
\affiliation{\iNANO}
\affiliation{\IFA}
\author{Brian Julsgaard}
\affiliation{\iNANO}
\affiliation{\IFA}
\email{brianj@phys.au.dk}

\date{\today}

\begin{abstract}
  Tin-containing nanocrystals, embedded in silicon, have
  been fabricated by growing an epitaxial layer of
  Si$_{1-x-y}$Sn$_x$C$_y$, where $x = 1.6$ \% and $y = 0.04$ \% on a
  silicon substrate, followed by annealing at various temperatures
  ranging from $650\,^{\circ}$C to $900\,^{\circ}$C. The nanocrystal
  density and average diameters are determined by scanning
  transmission-electron microscopy to $\approx
  10^{17}\:\mathrm{cm^{-3}}$ and $\approx 5$ nm,
  respectively. Photoluminescence spectroscopy demonstrates that the
  light emission is very pronounced for samples annealed at
  $725\,^{\circ}$C, and Rutherford back-scattering spectrometry shows
  that the nanocrystals are predominantly in the diamond-structured
  phase at this particular annealing temperature. The origin of the
  light emission is discussed. 
\end{abstract}

\maketitle

It is a long-lasting desire to combine electrical and optical
functionality in silicon (Si) \cite{Soref.ProcIEEE.81.1687(1993)}, the
achievement of which is hindered mainly by the indirect band gap of
Si. In order that light can be manipulated and trapped inside Si, the
photon energy must be well below the Si band-gap energy, which has led
researchers to investigate light emission from various combinations of
the other group-IV elements, carbon (C), germanium (Ge), and tin
(Sn). One possible route is to use Ge, which is also an
indirect-band-gap material, but the direct band gap can be approached
by strain engineering \cite{ElKurdi.ApplPhysLett.96.041909(2010)}. The
combination of tensile strain and high doping has led to the
demonstration of lasing in Ge \cite{Liu.OptLett.35.679(2010)}. Another
line of research includes Sn in the material, and in addition to the
strain conditions, the capability of efficient light emission now also
depends on the relative composition of Ge and Sn
\cite{Soref.JMaterRes.22.3281(2007),
  Ghetmiri.ApplPhysLett.105.151109(2014)}. A proper choice of
parameters has recently led to the demonstration of light-emitting
diodes \cite{Oehme.IEEEPhotonTechLett.26.187(2014)} and lasing
\cite{Wirths.NaturePhotonics.9.88(2015)} in GeSn alloys. In general,
it is possible to combine C, Si, Ge, and Sn in epitaxial film-growth
processes due to strain symmetrization
\cite{Guarin.ApplPhysLett.68.3608(1996)}, and the accessible parameter
space becomes very large. For instance, a broad-band luminescence peak
at 0.77 eV has been observed from SiSn films
\cite{Khan.ApplPhysLett.68.3105(1996)}, and this emission can be
enhanced by adding a small amount of carbon
\cite{Wright.MatResSocSympProc.533.327(1998)}. In the present work, we
follow yet another strategy: embedding nanocrystals  
containing   Sn inside crystalline
silicon. Diamond-structured $\alpha$-Sn is a direct-band-gap material
in bulk form, and the optical transition is predicted to remain strong
in the nanocrystal regime \cite{Jensen.PhysStatSolC.8.1002(2011)}. In
general, light emission is also possible from nanocrystals of
indirect-band-gap materials due to the break-down of the
momentum-conservation rule \cite{Kovalev.PhysRevLett.81.2803(1998)},
and, in addition, size control enables tuning of the emission
energy. In the longer perspective, an efficient nano-sized light
emitter might work as a single-photon source
\cite{Santori.Nature.419.594(2002)} in Si. The successful embedding of
Sn nanocrystals in Si has been realized by a number of research groups
\cite{Fyhn.PhysRevB.60.5770(1999), Ragan.MatSciEngB.87.204(2001),
  Karim.OptMaterials.27.836(2005), Tonkikh.JCrysGro(2015),
  Arslan.Science(2005)}, but studies of light emission is very
limited. Recently, new spectral features were reported in the infrared
region from Sn nanocrystals in SiGe quantum wells
\cite{Tonkikh.JCrysGro(2015)}, and previously
\cite{Karim.OptMaterials.27.836(2005)} a broad luminescence peak
around 0.85 eV was observed but not investigated in further detail. In
this work, we obtain a similar luminescence peak as in the latter
reference, and we demonstrate that its strength is much
enhanced when the nanocrystals are formed in the diamond-structured
phase.

\begin{figure}[t]
  \centering
  \includegraphics[width=\linewidth]{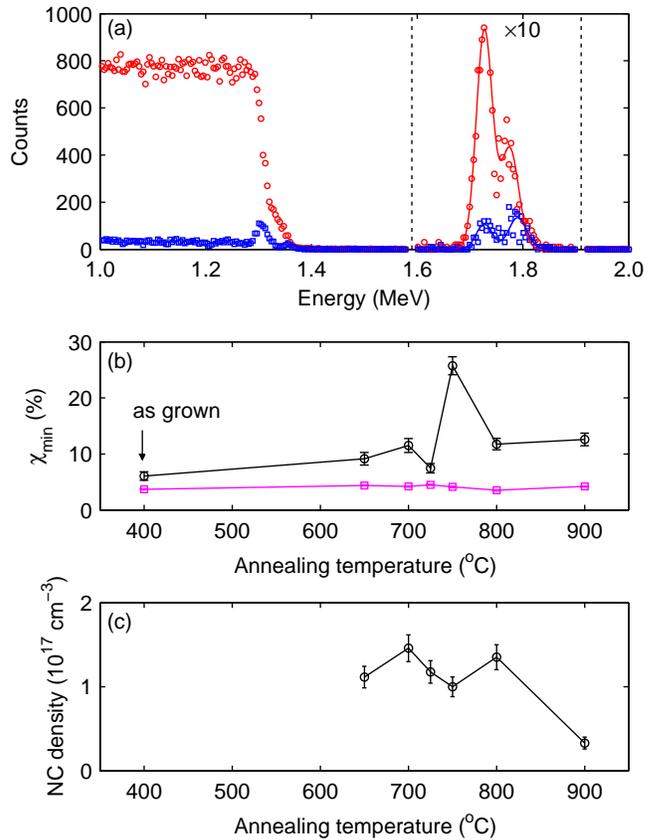}
  \caption{(a) The measured Rutherford-back-scattering spectrum for
    random orientation (red circles) and under channeling conditions
    (blue squares) for the sample annealed at $700\,^{\circ}$C. Solid
    lines are curve fits with two Gaussian peaks. Panel (b) shows
    $\chi_{\mathrm{min,Sn}}$ (black circles) and
    $\chi_{\mathrm{min,Si}}$ (magenta squares), as a function of the
    annealing temperature. Panel (c) shows, as a function of annealing
    temperature, the nanocrystal (NC) density determined by STEM.}
  \label{fig:RBS}
\end{figure}

The Sn-nanocrystal samples were prepared by molecular-beam epitaxy
(MBE) on a Si(100) wafer (75-125 $\Omega \cdot$cm). A 100 nm thick Si
buffer layer was grown followed by a 30 nm layer composed of Si, Sn
and C. The composite layer was co-deposited at $200\,^{\circ}$C at a
growth rate of 0.3 \r{A}/s and with ratios determined by
secondary-ion-mass spectrometry (SIMS) to 1.6 \% Sn and 0.04 \%
C.   The small amount of C is added in order to reduce the
strain and avoid dislocations
\cite{Gaiduk.ApplPhysLett.104.231903(2014)}. Finally, a 50
nm Si capping layer was grown at a temperature of
$400\,^{\circ}$C. Post-growth annealing was carried out in an
$\mathrm{N}_2$ atmosphere at temperatures varying from
$650\,^{\circ}$C to $900\,^{\circ}$C for 20 min.~in order to form the
nanocrystals.
\begin{figure*}[t]
\centering
 \includegraphics[width=\linewidth]{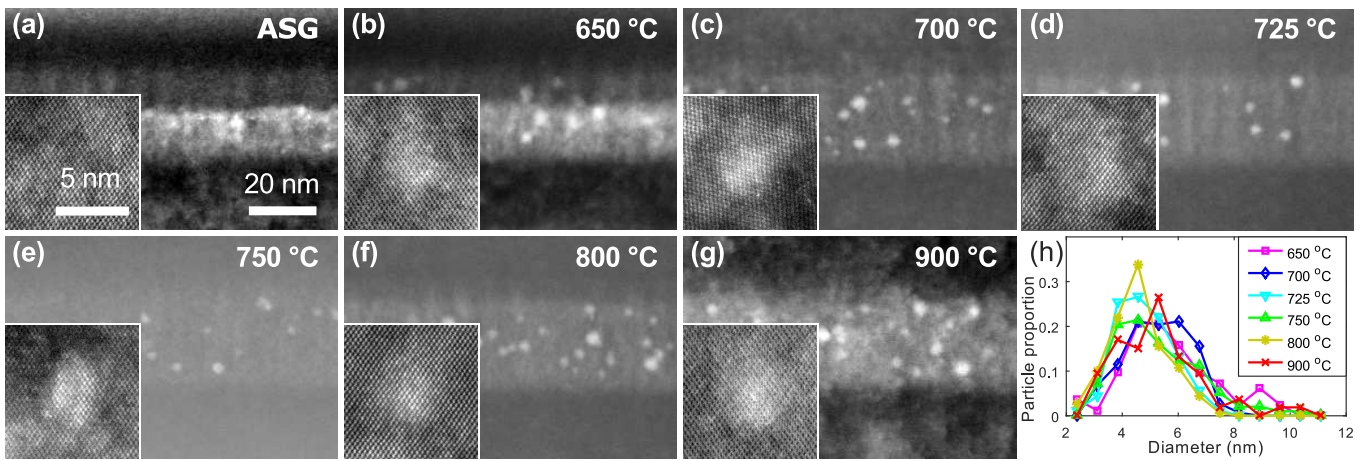}
 \caption{STEM images of SiSnC layers as-grown (ASG) and with
   post-annealing. The annealing temperatures are indicated in the
   images. Insets show high-resolution STEM images of the nanocrystals
   for the respective samples. Panel (h) shows histograms of the
   diameter distribution for the nanocrystals based on various STEM
   images. The samples have different thickness and the apparent
   nanocrystal density is therefore not representative.}
  \label{fig:STEM}
\end{figure*}
Cross-sectional samples, thickness of electron transparency, were
prepared for investigations by scanning transmission-electron
microscopy (STEM) using a focused-ion-beam (FIB) sputtering system
(FEI Versa 3D). The samples were investigated using a Talos A
microscope from FEI, with a high-angle annular dark-field (HAADF)
detector. Typical images for different samples, shown in
Fig.~\ref{fig:STEM}, demonstrate that nanocrystals are formed at all
the investigated annealing temperatures with only minor variations in
the size distributions, see Fig.~\ref{fig:STEM}(h). We measure the
thickness of each cross-section STEM sample by electron-energy-loss
spectroscopy (EELS), which in turn allows us to determine the volume
density of nanocrystals, see Fig.~\ref{fig:RBS}(c). The
high-resolution images (insets in Fig.~\ref{fig:STEM}) were obtained
by aligning the STEM electron beam along the $[110]$ direction and
forming the image contrast by inelastic Rutherford scattering, i.e.~by
variations in atomic number. Hence, each spot in the image corresponds
to a string of atoms, and the bright spots include Sn atoms on the
string.   Throughout the sample no change in lattice
parameters is seen between the nanocrystals and the background
material, and the boundary appears completely coherent.

The setup for optical investigations is identical to the one used in
Ref.~\onlinecite{Julsgaard.Nanotechnology.22.435401(2011)}. The
samples were excited by a short laser pulse of 100 fs duration and
wavelength 400 nm. The light emitted from the samples was dispersed by
a monochromator and measured by a photo-multiplier tube. The sample
temperature during optical investigations, not to be mistaken for the
annealing temperature, could be varied from room temperature down to
16 K.
\begin{figure}[t]
\centering
  \includegraphics[width=8.5cm]{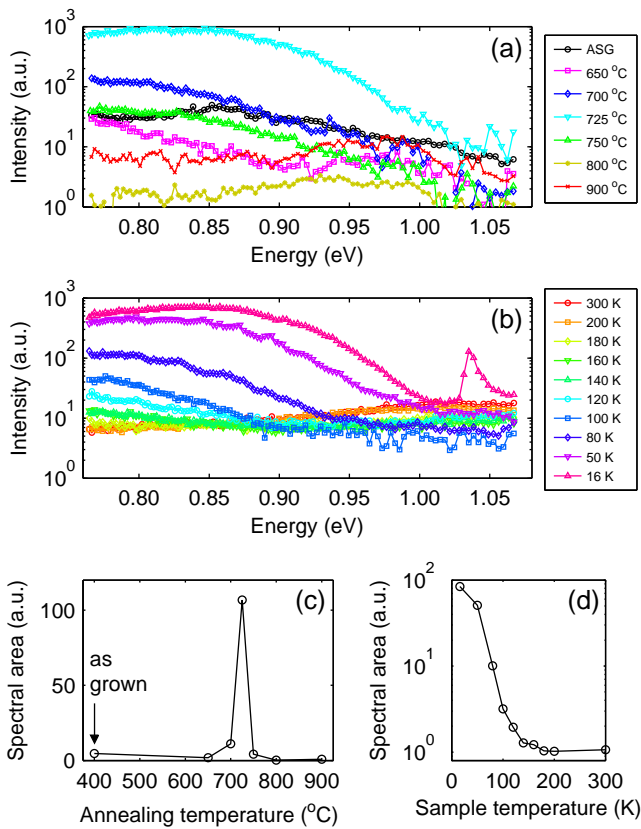}
  \caption{(a) Time-integrated ($0 \le t \le 600\:{\mu}\mathrm{s}$) PL
    spectra  (please note the log scale on the vertical
    axis), obtained from samples annealed at different
    temperatures and excited by a pump fluence of $\approx 8\cdot
    10^{-6}\:\mathrm{J/cm^2}$ at a sample temperature of 16 K. The
    area below each curve, for emission energies $E \le 0.90$ eV, is
    plotted in panel (c), showing a pronounced peak at
    $725\,^{\circ}$C. Panel (b) shows the time-integrated ($0 \le t
    \le 10\:{\mu}\mathrm{s}$) PL spectra for the sample annealed at
    $725\,^{\circ}$C, for various sample temperatures using a pump
    fluence of $\approx 1.5\cdot 10^{-4}\:\mathrm{J/cm^2}$. The area
    below the curves, for $E \le 0.90$ eV, is plotted in panel (d).}
  \label{fig:Optical}
\end{figure}

The results of the optical investigations are summarized in
Fig.~\ref{fig:Optical}. In Fig.~\ref{fig:Optical}(a) the
photoluminescence (PL) spectra are shown for various annealing
temperatures and for the as-grown (ASG) sample. We note that all
samples emit some light, consistent with previous studies
\cite{Khan.ApplPhysLett.68.3105(1996),
  Wright.MatResSocSympProc.533.327(1998)} of as-grown SiSnC
films.   There is, however, a remarkably strong and rather
broad luminescence feature centered around 0.82 eV from the sample
annealed at $725\,^{\circ}$C. Even though the center of these features
depends on the annealing temperature and parts of the luminescence lie
below the detection limit of 0.76 eV, it is evidently the height of
each feature that is main responsible for the increased
luminescence. To illustrate the overall behavior of the luminescence as
a function of annealing temperature, we have calculated the
area under each spectral curve, from 0.76 eV to 0.90 eV, and plotted
these areas in Fig.~\ref{fig:Optical}(c). The strongest luminescent
sample (annealed at $725\,^{\circ}$C) has also been studied at various
sample temperatures as shown by the spectra in
Fig.~\ref{fig:Optical}(b) and the spectral areas in
Fig.~\ref{fig:Optical}(d). The luminescent peak persists from low
temperatures up to roughly 150 K.

We complement the structural STEM analysis shown in
Fig.~\ref{fig:STEM} by Rutherford-back-scattering spectrometry (RBS),
using 2 MeV helium (He) ions, see Fig.~\ref{fig:RBS}. In
Fig.~\ref{fig:RBS}(a) the broad feature below 1.4 MeV arises from Si,
whereas the signal (multiplied by 10) between the vertical dashed
lines originates from Sn. The left Sn peak, centered at 1.73 MeV,
corresponds to back scattering from buried Sn atoms, and the adjacent
shoulder towards higher energy originates from Sn atoms at the sample
surface. When the incident He ions are aligned to channel along the
$[001]$ direction, the probability of a large-angle back-scattering
event will be strongly reduced. This is seen, e.g., as a reduction of
the Si yield to $\chi_{\mathrm{min,Si}} = 4$ percent of the random
yield under such channeling conditions, where $\chi_{\mathrm{min,Si}}$
was calculated as the channeling-to-random area under the RBS spectra
in the range 1.0 MeV to 1.2 MeV. Likewise, a reduction of the peak
area at 1.73 MeV (determined from a Gaussian fit) will indicate that
buried Sn atoms are also suppressed by the channeling conditions and
hence occupy predominately substitutional or near-substitutional
sites. Calculating $\chi_{\mathrm{min,Sn}}$ as the
channeling-to-random ratio of this peak area, shown by the black
circles in Fig.~\ref{fig:RBS}(b), we clearly see variations in this
area ratio, i.e.~in the degree to which the Sn atoms are coherent with
the surrounding Si lattice. A low value of $\chi_{\mathrm{min,Sn}} =
(7.5 \pm 0.8)$ \% is observed at $725\,^{\circ}$C, corresponding to a
substitutional fraction \cite{Feldman_Mayer_Picraux} of $S =
(1-\chi_{\mathrm{min,Sn}})/(1-\chi_{\mathrm{min,Si}}) = (96.9 \pm
0.9)$ \%, which is followed by a rapid increase to
$\chi_{\mathrm{min,Sn}} = (25.7 \pm 1.6)$ \% at $750\,^{\circ}$C,
corresponding to a substitutional fraction of $S = (77.5 \pm 1.7)$
\%. In comparison, we measure for the as-grown sample the values
$\chi_{\mathrm{min,Sn}} = (6.1 \pm 0.8)$ \% and $S = (97.5 \pm 0.8)$
\%.

The STEM data demonstrate that the variations in the nanocrystal size
and number density are rather limited, taking the quite broad range of
annealing temperatures into account. This is consistent with the fact
that a Sn atom requires a vacancy to assist its diffusion
\cite{Kringhoj.PhysRevB.56.6396(1997)}. These vacancies are most
likely present from the initial low-temperature MBE growth process
\cite{Gossmann.ApplPhysLett.61.540(1992)} and available during both
the first Sn-Sn segregations steps
\cite{Fanciulli.PhysRevB.61.2657(2000)} and the following coarse
precipitation stage, which must take place at a temperature below
$650\,^{\circ}$C. In other words, for the MBE-grown samples, the Sn
precipitation does not depend on thermally generated vacancies, which
explains the limited variations in the nanocrystal size distribution
on annealing temperature. Next, the different investigated annealing
temperatures must lead to further maturing stages of the Sn
precipitates and eventually define the conditions for the finer
variations in the resulting nanocrystal structure upon cool-down to
room temperature. These variations can be seen from the RBS results in
Fig.~\ref{fig:RBS}(b), in particular in the range $700\,^{\circ}$C to
$800\,^{\circ}$C. Evidently, for an annealing temperature of
$750\,^{\circ}$C a significant fraction of Sn atoms ends up in
non-substitutional sites. It is clear from Fig.~\ref{fig:STEM} that
some of the Sn atoms always remain in solution within the original
SiSnC layer (seen as a bright layer of thickness 30 nm in the
images). However, a significant fraction of these dissolved Sn atoms
cannot be located at non-substitutional sites. If they were, there
would be a high probability for low-angle scattering events for all He
ions passing through the Sn-rich layer. This would lead to
de-channeling and accordingly a large $\chi_{\mathrm{min,Si}}$ for
back-scattering from deep Si below the SiSnC layer
\cite{Feldman_Mayer_Picraux}. This is, however, not observed; the
magenta squares in Fig.~\ref{fig:RBS}(b), representing these deep Si
atoms, remain at a low value around 4 \% for the sample annealed at
$750\,^{\circ}$C. Hence, the non-substitutional Sn must be located
primarily within the nanocrystals. These Sn atoms will still cause
de-channeling and should potentially raise the
$\chi_{\mathrm{min,Si}}$-level for the deep Si atoms. But since the
nanocrystal volume density is $\approx 10^{17}\:\mathrm{cm^{-3}}$, the
area density (calculated using the 30 nm layer thickness) must be
$\approx 3\cdot 10^{11}\:\mathrm{cm^{-2}}$, and the $\approx 5$ nm
nanocrystals will then take up only $\approx 6$ \% of the projected
area. In consequence, $\approx 94$ \% of the He ions will never risk a
low-angle scattering event, and in practice no de-channeling is
observed in the Si part of the RBS spectrum.

  Before we start interpreting the experimental results, it
should be stressed that the amount of Sn (and C) in the nanocrystals is
unknown. The low solid solubility of Sn in Si around 0.1 \%
\cite{Trumbore.BellSysTechJ.39.205(1960)} suggests that the
nanocrystals consist of pure Sn. However, even a small amount of C is
known to have an impact on the Sn segregation process
\cite{Gaiduk.ApplPhysLett.104.231903(2014)}, and thus possibly also on
the resulting nanocrystal composition. A recent study has also
indicated that the nanocrystals can exist as an SiSn alloy
\cite{Tonkikh.JCrysGro(2015)}. Now, for the sample
annealed at $750\,^{\circ}$C the question arises as to how the
nanocrystals contain, on the one hand, non-substitutional Sn (as
determined by RBS) and still, on the other hand, give rise to the
coherent high-resolution images from STEM. A possible answer could be,
  under the assumption of a pure Sn nanocrystal,  
that the apparent coherence holds for an $\alpha$-Sn shell and at the
same time the core might be something else, possibly $\beta$-Sn, not
aligned to the $[110]$ direction and contributing a flat contrast
level to the high-resolution STEM images. Along the same lines of
thought, the sample annealed at $725\,^{\circ}$C will be most likely
to contain pure or high-quality $\alpha$-Sn nanocrystals due to the
low value of $\chi_{\mathrm{min, Sn}}$. It is remarkable that the
variations in the nanocrystal structure, evidenced by
Fig.~\ref{fig:RBS}(b), and the associated variations in
photoluminescence intensity, evidenced by Fig.~\ref{fig:Optical}(c),
occur in such a narrow annealing temperature range. Nevertheless, the
above   suggestions   are consistent with previous
findings from M\"ossbauer spectroscopy
\cite{Ridder.MatSciSemProc(2000)}, which also demonstrated that only a
narrow annealing-temperature window leads to formation of pure
$\alpha$-Sn nanocrystals whereas higher annealing temperatures lead to
an additional formation of $\beta$-Sn. Although for significantly
larger nanocrystals, it was also found in
Ref.~\onlinecite{Fyhn.PhysRevB.60.5770(1999)} that, upon cooling,
molten Sn precipitates first into an $\alpha$-Sn formation, coherent
with the surrounding Si, followed by a $\beta$-Sn formation due to a
rise of pressure. The above discussion supports the suggestion that
the sharp rise in $\chi_{\mathrm{min,Sn}}$, at the annealing
temperature of $750\,^{\circ}$C, could have its origin in the onset of
a $\beta$-Sn core. It is more unclear why the $\chi_{\mathrm{min,Sn}}$
decreases again at the annealing temperature of $800\,^{\circ}$C, but
it could be related to part of the Sn going back into solution, which
is weakly indicated by the results of
Ref.~\onlinecite{Ridder.MatSciSemProc(2000)} and by the decreasing NC
density towards $900\,^{\circ}$C annealing temperature.

In the bulk form and at atmospheric pressure, the melting point of
Si$_{1-x}$Sn$_x$ is known \cite{Olesinski1984} to depend strongly on
$x$, and the corresponding melting point in nanocrystalline form is
unknown. This adds further complexity to the nanocrystal formation
process and its underlying kinetics,   if we assume that the
Sn concentration in the nanocrystals is less than unity. Even under this assumption, however, our work concludes that the luminescence is strongest when the
crystalline quality of the diamond-structured nanocrystals is
highest. 

Based on the optical and structural observations, one suggestion is
that the light emission originates from the $\alpha$-Sn or  
Sn-rich   nanocrystals. This conjecture is consistent with
the temperature-dependent PL observations, shown in
Fig.~\ref{fig:Optical}(b,d), since the absence of light emission at
high sample temperatures could reflect thermal excitation of electrons
and holes into the surrounding bulk conduction and valence bands. The
center emission energy of 0.82 eV is consistent with
Ref.~\onlinecite{Jensen.PhysStatSolC.8.1002(2011)} for a 5 nm
$\alpha$-Sn nanocrystal subjected to moderate compressive strain. Our
sample, annealed at $725\,^{\circ}$C and presumably with a crystal
structure coherent with the surrounding Si, is probably subjected to a
quite high compressive strain leading to a larger predicted emission
energy. However, the calculations in
Ref.~\onlinecite{Jensen.PhysStatSolC.8.1002(2011)} assumed the
nanocrystal to be surrounded by vacuum, while our Si environment may
allow for a less confined wave function and accordingly smaller
confinement and emission energies. We emphasize that light emission
from dislocations \cite{Drozdov.JETPLett(1976)} in Si is unlikely,
since no extended dislocations were visible in STEM.
It cannot be excluded, though, that the luminescence could originate from point defects at the interface between nanocrystals and the surrounding material.  

In conclusion, we have fabricated Sn-containing nanocrystals inside crystalline Si
and studied their ability to emit light by photoluminescence
spectroscopy. A pronounced enhancement of light emission, in a broad
band centered at 0.82 eV, is found for the sample annealed at
$725\,^{\circ}$C, which is
 also the sample where the nanocrystals are
most coherent with the surrounding Si lattice, i.e.~containing Sn in a
high-quality diamond-structured phase. Our results are consistent with the interpretation that light originates from the nanocrystals, although other emission mechanisms cannot be excluded.

This work was supported by the Villum Foundation. We are grateful to
Bjarke R.~Jeppesen for assistance with the STEM sample preparation.

%

%

\end{document}